# Towards final characterisation and performance of the GCT prototype telescope structure for the Cherenkov Telescope Array

**O. Le Blanc, G. Fasola, J.L. Dournaux, L. Dangeon, V. Hocdé, J.P. Amans, J.M. Huet, I. Jégouzo, P. Laporte, H. Sol, F. De Frondat, C. Perennes**[*]

*LUTH and GEPI - Observatoire de Paris, PSL Research University, CNRS*
*5, place Jules Janssen - 92190 Meudon, France*
*E-mail:* oriane.le-blanc@obspm.fr

**J. Gironnet, G. Buchholtz, A. Abchiche for the CTA GCT Project**

*DT-INSU, CNRS, University Paris Diderot*
*1 place Aristide Briand - 92190 Meudon, France*

*http://www.cta-observatory.org*

The Gamma-ray Cherenkov Telescope (GCT) is an innovative dual-mirror solution proposed for the Small-Size-Telescopes of the future Cherenkov Telescope Array (CTA), capable of imaging the showers induced by cosmic gamma-rays with energies from a few TeV up to 300 TeV. The Schwarzschild-Couder design on which the telescope optical design is based makes possible the construction of a fast telescope (primary mirror diameter 4 m, focal length 2.3 m) with a plate scale well matched to compact photosensors, such as multi-anode or silicon photomultipliers (MAPMs and SiPMs, respectively) for the camera. The prototype GCT on Meudon's site of the Observatoire de Paris saw first Cherenkov light from air showers in November 2015, using an MAPM-based camera. In this contribution, we firstly report on the prototype GCT telescope's performance during its assessment phase. Secondly, we present the telescope configuration during a campaign of observations held in spring 2017. Finally, we describe studies of the telescope structure, such as the pointing and tracking performance.

1.  Introduction

The Cherenkov Telescope Array (CTA) [1] is the next major international instrument in very-high energy astrophysics (20 GeV-300 TeV), providing a sensitivity of one order of magnitude better than present instruments as well as a significant improvements in angular resolution. Given the wide energy range to be covered, three different size classes of Imaging Air Cherenkov Telescopes (IACTs) designed for different energy ranges are therefore proposed and will be located in both the Northern and the Southern hemispheres to observe the whole sky. Among them, the Small Size class of Telescopes (SSTs) is devoted to the highest energy region from a few Tev to 300 TeV. About 70 of these telescopes are foreseen in the southern array to be constructed in northern Chile, between ESO's Paranal and Cerro Armazones sites [2].

The Gamma-ray Cherenkov Telescope (GCT) is one of three proposed solutions for the SSTs. Designed and built by an Australian-Dutch-French-German-Japanese-UK-US consortium, the GCT uses an innovative dual-mirror Schwarzschild-Couder (SC) optical design [3] in order to increase the field of view (FoV) while reducing the required camera size. The GCT prototype telescope structure has been installed at the Meudon site of the Observatoire de Paris in France and has been operated since November 2015. Two prototype cameras are under development for GCT: one using Multi-Anode Photomultiplier Tubes (MAPMTs) known as CHEC-M, and a second one using Silicon Photomultiplier Tubes (SiPMTs) known as CHEC-S [4]. One year and a half after recording CTA's first ever Cherenkov light, CHEC-M was shipped back to Meudon for a commissioning and observing campaign.

In the following, the GCT prototype telescope performance as well as the latest campaign of observation in Meudon are briefly described. A further section focuses on the developments made in terms of design validation or integration during this last year from the project status reported in Dournaux et al. [5].

2.  Telescope structure overview

Design studies of the mechanical structure started in 2011 at the Observatoire de Paris in France and have been led with a desire (i) to provide a mechanical structure as simple and as light as possible, (ii) to ease the mounting and maintenance phase and (iii) to decrease manufacturing costs by using commercial-of-the-shelf (COTS) modules and similar systems in the telescope. The integration of the mechanical structure from pre-assembled and pre-set subsystems was held on the Meudon site of the Observatoire de Paris in 2015. The majority of the structural frame is made of standard steel E355. The telescope structure itself is comprised of four main subsystems [6, 7] listed hereafter:

- The tower. It is a COTS tube equipped with two flanges. It provides a mechanical interface between the telescope and the foundation and also supports the mass of the telescope.
- The alt-azimuthal mechanical structure (AAS). The GCT is equipped with an alt-azimuth mount. The AAS is a simple and modular structure, which consists of the fork, the drive systems and the bosshead. The designs of the AAS drives for the elevation and azimuth axes are similar and are made up of one slew bearing with an encoder and a motor shaft formed of one worm gear with two motors and encoders.

- The optical support structure (OSS) and the counterweight (CW). The OSS consists of the mast (COTS tubes), the dish of the primary mirror (M1), the top dish that supports the secondary mirror (M2), the bottom dish that links the OSS to the bosshead, and the camera removal system.
- The optical elements. The GCT is equipped with two aspherical concave mirrors: M1 and M2. Lightweight aluminium mirrors are used for the GCT prototype [8]. The M1 mirror is tessellated and made of six petals. Only two petals are currently in place on the prototype, the other four are dummies with a similar surface area and mass to those of the real petals. The secondary mirror is monolithic. The camera is developed by the CHEC consortium [4].

The complete and assembled structure equipped with the CHEC-M camera is shown in Fig. 1. The final design of the telescope has a total mass of about 8 tons.

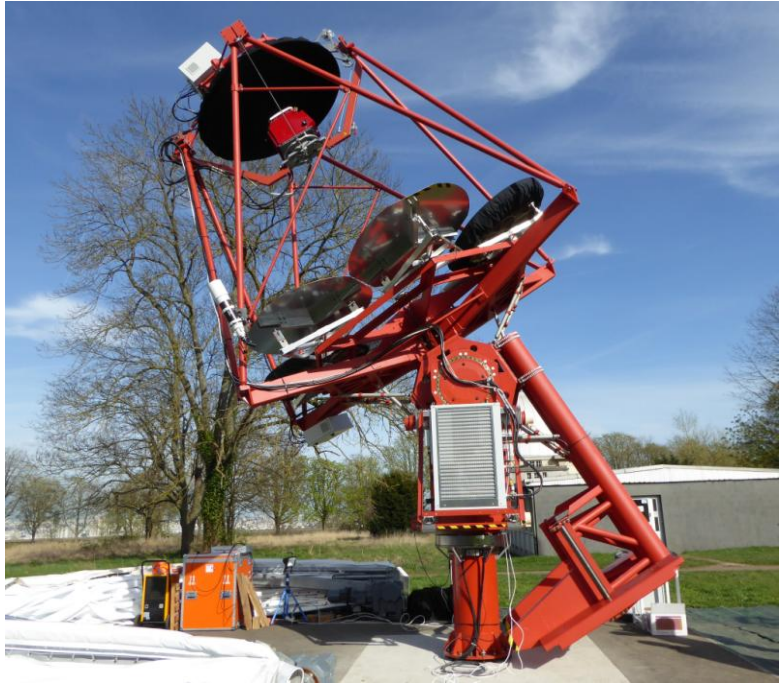

**Figure 1:** *The GCT prototype equipped with its camera (CHEC-M) in March 2017.*

The Telescope Control System's (TCS) main purpose is to drive the telescope's axes for tracking and pointing [5, 9]. To achieve this, three embedded cabinets connected to a power supply and network (Internet and fieldbus) are implemented on the telescope. The hardware for slow-control and safety systems are COTS modules from Beckhoff. To control the motion of the prototype, the TCS software is distributed between the hardware (drives and motors provided by ETEL) on board the telescope and the workstation of the control room (see Fig. 2).

To achieve pointing, the equatorial coordinates of a source have to be converted into altitude-azimuth – or horizontal – coordinates, taking into account the observatory location and the date and time. For tracking, the conversions have to be computed for real-time operation to allow accurate scientific observations. The pointing and tracking software (written in C/C++) were integrated in early 2017 in the prototype embedded PC, in accordance with the smart telescope design philosophy. The SOFA library, referenced by the International Astronomical Union, is used for astronomic calculations in compliance with the CTA recommendations.

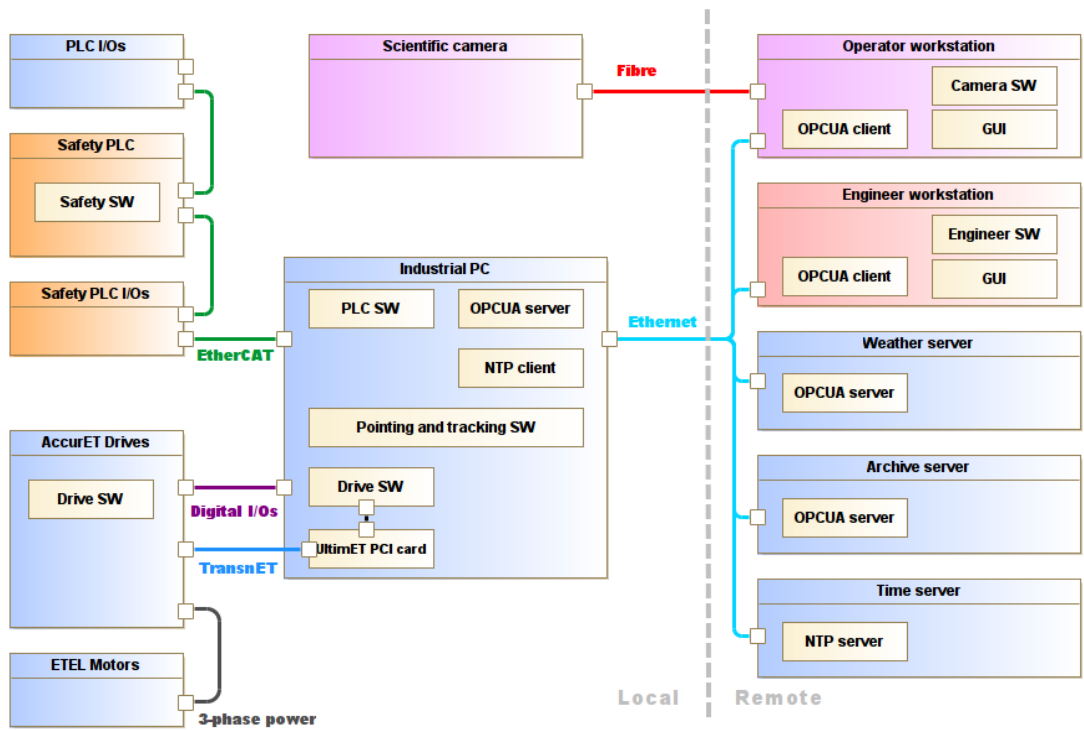

**Figure 2:** *An overview of both the software and the communication layout of the TCS. There are only two physical links between the telescope and the control room for communication purposes. An EtherCAT fieldbus is used for safety and local automation. The logical connectivity is managed by OPC-UA clients and servers. The network devices, such as switches, are not shown.*

Following the Design Verification Document and the conformity matrix, 246 GCT technical specifications were defined in order to verify that the prototype meets the CTA requirements. Each specification has to be checked by one or more verification methodologies: analysis, certification, demonstration, inspection and testing [5].

The GCT prototype has therefore undergone intensive analyses and testing in 2016 and 2017 and no technological barrier or specific risks has been identified. A laser placed at 75 m from the telescope has been used, as detailed in [8], to align the M1 petal equipped with manual actuators. To optimise the pointing precision, an ATIK monochromatic camera has been installed at the focal surface in place of the scientific camera (in a CHEC-S dummy housing reproducing the actual weight and inertia of the scientific camera) to collect enough pointing data of stars in order to obtain a good model of the telescope geometry.

The most constraining tests concern the movement of the telescope and specifically, the maximum possible speed, the efficiency of the system and the emergency stop. Quantitative measurements of the motion characteristics of the azimuth and the elevation axes have been performed and are summarized in Table 1. To date, about 80 % of the specifications have been processed and passed.

| Designation | Specification | Test Result |
|---|---|---|
| Azimuth range | 510 ° | 523.8 ° |
| Elevation range (tracking range) | 91 ° (89.2 °) | 90.7 ° (89.6 °) |
| Max azimuth pointing speed | 5 °/s | 6.0 °/s |
| Max elevation pointing speed | 2 °/s | 2.2 °/s |
| Max azimuth tracking speed | 0.3 °/s | 0.38 °/s |
| Max elevation tracking speed | 0.0039 °/s | 0.0039 °/s |
| Lifetime of the mechanical structure | 30 years | 42.4 years |
| Emergency stop button action | < 1 s | < 1 s |

**Table 1:** *Summary of mechanical and drives performances compared to the CTA specifications for SSTs.*

## 3. Observing campaign of spring 2017

The main goals of the campaign were end-to-end tests of the safety and operation procedures for the camera and telescope. For the first time, both camera and telescope prototypes were operated remotely and in concert from the nearby control room in a reliable manner (the telescope's remote and embedded PCs and the camera control PC all being accurately synchronized with the Network Time Protocol server of the Observatoire de Paris). This included monitoring of all housekeeping data, visual checks using a webcam, drives temperatures, telescope position, power consumption, weather conditions, etc. It was also the first test of the camera's improved safety and control system.

The safety of the operating personnel was ensured thanks to hazard warnings on and around the telescope site, as well as an audible warning sounding when the telescope is slewing. All Observatory campaign participants were trained to act as telescope operators during observing nights and a "Telescope Operator User Manual & Safety rules" for the prototype was issued.

During the observing campaign, the GCT camera acquired thousands of events that are being analysed. Two examples of on-sky images are shown in Fig. 3 (right), for details on the image analysis please refer to the proceedings by R. White and H. Schoorlemmer [4] as well as H. Sol et al. [10] for the CTA GCT Project in this issue.

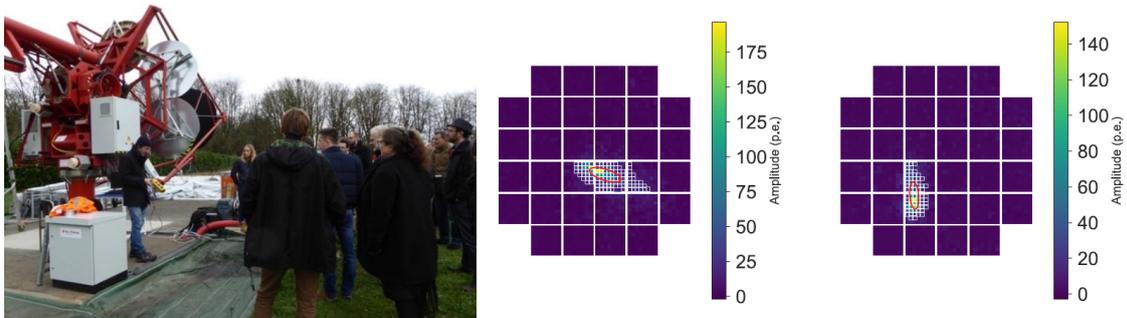

**Figure 3:** *On the left, safety brief during the observing campaign (03/2017). On the right, a selection of two Cherenkov shower events detected by CHEC-M on the GCT prototype in Meudon. The calibrated image intensity in pe is shown for each camera pixel.*

## 4. Performance

### 4.1 Mirrors quality

The GCT prototype mirrors conformity to CTA requirements was confirmed by a measurement campaign using an X-rite spectrophotometer (bandwidth considered: 400 to 550 nm) in accordance with the studies reported in [8]. These results are presented in Table 2.

| Mirrors | Total reflectivity | Scattered component |
|---|---|---|
| M1 | 89 % (+/- 0.5 %) | 7.7 % (+/- 2 %) |
| M2 | 87 % (+/- 1 %) | 10.3 % (+/- 1.5 %) |

**Table 2:** *Mean values of the reflectivity of the GCT prototype metallic mirrors.*

### 4.2 Drive performance

Concerning the drive's power consumption, Fig. 4 shows the apparent power measured at the main power supply when the azimuth axis is in motion at different velocities with all other subsystems switched off. The mean apparent power required for azimuth slewing is 200 VA with a peak at about 230 VA (see [10] for figure). This test result confirms that the technical choices made for the AAS drives fulfil easily the current CTA specification (maximum power available in the CTA power grid is about 10 kW).

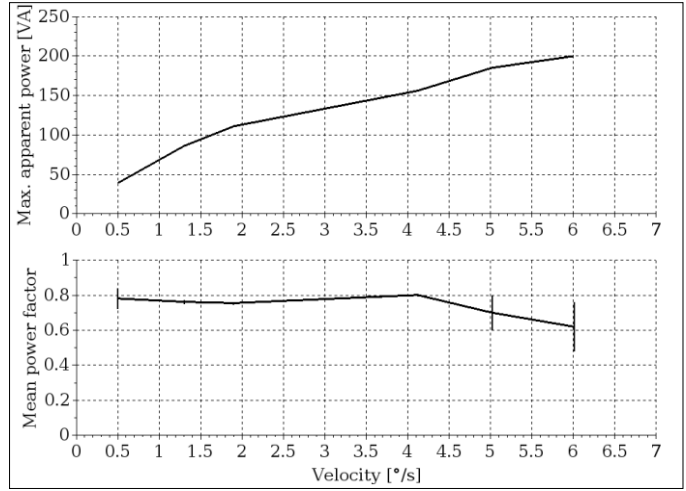

**Figure 4:** *Maximum required apparent power and power factor versus velocity of the azimuth axis in standard observing conditions.*

### 4.3 Tracking performance

We show the pre-calibration results in Table 3 of the tracking error between the commanded trajectory and the telescope position (alt-azimuthal) read by the axes absolute angle encoders fitted in the slew bearings during a half-an-hour tracking. The small tracking errors before any optimisation of the pointing are very encouraging.

|  | Azimuth trajectory (°) | Elevation trajectory (°) |
|---|---|---|
| Maximum discrepancy | -0.066 | -0.014 |
| Mean discrepancy | -0.063 | -0.013 |
| Standard deviation | 0.001 | 0.001 |

**Table 3:** *Maximum, mean and standard deviation in degrees of the discrepancies observed between the commanded trajectory and the telescope trajectory during a 30-minute tracking (before optimisation of the pointing accuracy).*

Figure 5 shows the recorded velocity of the azimuth axis while tracking at a high elevation pointing, near the zenith; this is where the constraints are the highest. The goal was to achieve a tracking at 89.2 ° in elevation, which results in a velocity of the azimuth axis of 0.27 °/s at 24° of latitude where the CTA southern site stands.

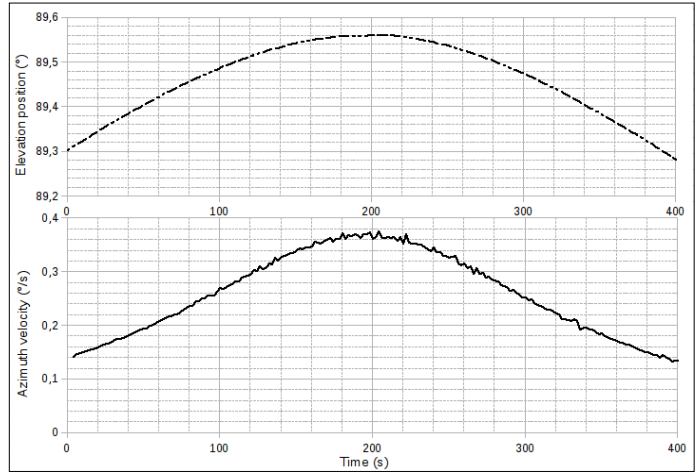

**Figure 5:** *Testing drive performance: velocity of the azimuth axis while tracking near the zenith. The required tracking velocity of the azimuth axis of 0.27 °/s is surpassed.*

### 4.4 Pointing performance

To enhance the pointing accuracy of the GCT prototype in any direction, many technical pointing runs were taken to gather data. We have recorded the position of the telescope along with the environmental conditions, while recording images of stars in the telescope FoV.

In order to take into account the pointing error due to the telescope geometry in the calculation of the alt-azimuthal coordinates, we have built a preliminary pointing model of the telescope using P. Wallace's TPOINT software [11]. In this preliminary model, we have considered only 7 parameters, i.e. AN, AW, IA, NPAE, IE, CA and TF.

The theoretical result is presented in Fig. 6: with a simple model and 30 samples, we achieve a factor of 20 improvement in the pointing precision of the telescope. This first result confirms that the telescope has a very good behaviour over the sky and that the theoretical pointing precision before applying any predictive model for the deformations due to gravity and wind, and before improvement using real time monitoring devices, exceeds expectations considering the CTA requirements (210 " in ideal observing conditions and 600 " in standard conditions). Henceforth, the telescope model needs to be refined by gathering more data with various observing environmental conditions.

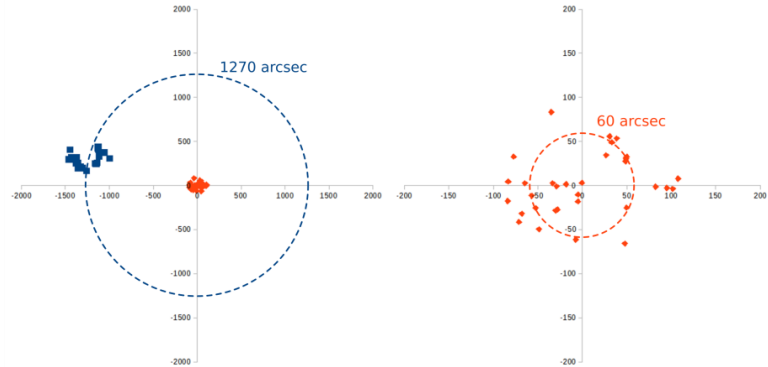

**Figure 6:** *On the left, the "scatter diagram" of the observed source image position on the CCD camera, before (blue) and after (orange) applying the first pointing model of the telescope. On the right, zoom in on the left diagram. The dotted circles stand for the mean values of each series.*

## 5. Conclusion and perspectives

The assessment phase of the GCT prototype is now coming to an end. The immediate next steps for the GCT telescope team will be the implementation of motorised actuators to optimise the optics alignment and conclude on its optical parameters. CHEC-M will remain in Paris, where the Observatory staff supported remotely by the camera team, will carry out routine operation. The focus will be on understanding the stability and the reliability of the systems.

Work is now in progress to update the detailed design plans for GCT-01, the first GCT telescope to be built in pre-production for the southern array of CTA, based on the knowledge and the expertise gained during the prototyping and assessment phases. The GCT consortium plans to build about 30 serial telescopes for the CTA Observatory, which will be provided from 2019 to 2022 in accordance with the southern CTA site construction plans.

**Acknowledgements**

This work has been done in the context of the CTA Consortium. We gratefully acknowledge financial support from the agencies and organizations listed here: http://www.cta-observatory.org/consortium_acknowledgments.